\documentstyle[preprint,aps]{revtex}

\begin{document}
\title
{Anomalous Dimension and Spatial Correlations in a Point-Island Model} 
\author{Ji Li, A.G.  Rojo, Leonard M. Sander}
\date{\today}
\draft
\address{H. M. Randall Laboratory of Physics
The University of Michigan, Ann Arbor, MI, 48109-1120}
\maketitle
\begin{abstract} 
  We examine  the island size distribution  function  and spatial 
correlation function
of a model for island growth in the submonolayer regime in both 1
and 2 dimensions. In our model the islands do not grow in shape,
and a fixed number of adatoms are added, nucleate, and are trapped
at islands as they diffuse. 

We study the cases of various critical island sizes $i$ for nucleation as a    
function of initial coverage. We found  anomalous scaling of the island
size distribution for large  $i$ .  Using scaling, random walk theory, a 
version of mean-field 
theory we obtain  a closed form for the spatial correlation function. Our 
analytic results are  verified by Monte Carlo simulations.

\vspace*{2in}
\begin{Large}
\end{Large}
\end{abstract}

\pagebreak 
Recently there has been much interest in  
island size distribution and morphology in
submonolayer epitaxial growth 
\cite{PRL1995Family,PRB1992Evans,EUROPHYS1991,PRB1993Evans,JFraceTang,Langmuir1996}. 
With the development of real-space
imaging methods such as scanning tunneling microscopy (STM) and atomic force
microscopy (AFM) ) and surface-sensitive diffraction methods (e.g. 
low-energy electron diffraction), several  experimental systems have been  
studied under conditions that favor irreversible aggregation and low island 
mobility and dissolution. Some examples are homoepitaxial systems
such as Fe/Fe(100) \cite{Fe/Fe(100)}  Ni/Ni(100) \cite{Ni/Ni(100)} Si/Si(100)\cite{Si/Si(100)}
Cu/Cu(100)\cite{Cu/Cu(100)} and 
heteroepitaxial systems such as Pb/Cu(001)\cite{Pb/Cu(001)}, Au/Ru(0001)\cite{Au/Ru(0001)} and
Ag/Si(111)\cite{Ag/Si(111)}.

In this paper we consider a simplified model in order to look at general
features of this type of growth. Our model is a point-island model in which we  ignore the island shape,   
and turn off the flux of adatoms. 
We put all our adatoms down at once and allow them to aggregate until the 
evolution stops.
We ignore all the complications which arise from shapes such as
dendritic or diffusion-limited-aggregation (DLA-like) fractal aggregates 
\cite{Au/Ru(0001),DLA}. 
We focus here on island size and island-island spatial correlation
scaling behavior. We find striking results in the limit of low coverage,
in particular a change in the scaling or the island-size distribution as we change 
the critical island size (defined below). We find that for large critical
island size, both the island size distribution ans the island-island 
correlation obey anomalous scaling.

In our model we start by randomly  distributing adatoms on a $d$-dimensional cubic 
lattice of size $L^d$. Each site is occupied with probability $\theta$. 
In principle the only natural length scale in this problem is $\theta^{-1/d}$, the 
initial mean distance
 between adatoms. Starting at time $t=0$, adatoms
are picked randomly  and diffuse by nearest-neighbor hopping. A stable island at site $x$
can be formed if $i$ adatoms meet simultaneously at $x$,  and any adatom stepping onto 
an existing stable island site will be absorbed   and becomes immobile. The size $s$ 
of an  island  is defined as the number of 
adatoms it has captured.  An island with size smaller than the  {\it critical size} $i$ 
can  dissociate without any energy barrier, while a stable island can not  dissociate  
and is immobile. The critical size  $i$ defined in this paper  differs by 
one from  the critical size usually defined 
in other literature. 
We have no incident flux of adatoms  in our model;   
at the end of the process only  stable islands  will survive.
One motivation for studying this model comes from an experiment \cite{MarkJohnson} recently performed 
where a 50-layer GaAs surface is quenched from a high temperature (600$^{\circ}$C) with the 
flux of adatoms shut off. However, our primary interest is in the model as 
an example of statistical physics far from equilibrium.


The first important quantity in the description of the asymptotic behavior of 
this model is the island-size distribution function $N_s$. This is the density
 of islands of size $s$. Assuming that there 
exists only one characteristic size in the problem which is the average island 
size $S$, one may guess a form for this function \cite{AmarFamLam}: $N_s(\theta)=Af(s/S)$,
where $f(u)$ is a scaling function which also depends on the critical 
island size $i$. Since there is no flux of atoms in this model, the 
number of atoms is conserved, hence $\theta = \sum_{s\geq 1} sN_s$.
 Approximating the sum by an integral, $\theta=AS^2\int_0^\infty f(u)udu $,  
 which implies $A \sim \theta/S^2$. Therefore 
one may write:
\begin{equation}
\label{Ns_Scale1}
N_s(\theta) \sim  \theta S^{-2}f(s/S)
\end{equation}
where the scaling function $f(u)$ satisfies $\int_0^\infty f(u)udu=1$. 
The total island density is given by 
 $N=\sum_{s\geq 2} N_s $ and the average
  island size is 
$
S=\sum_{s\geq 1} sN_s / \sum_{s\geq 1} N_s = \theta / N
$
where we have used the fact that $N_1=0$ at $t=\infty $.

We now assume that the average  island size S scales as 
\begin{equation}
\label{S_Scale1}
S \sim \theta^{-z}.
\end{equation}
Equation (\ref{Ns_Scale1}) may then be rewritten in the form
\begin{equation}
\label{Ns_Scale2}
N_s(\theta)=\theta^{1+2z}f(s \theta^z)
\end{equation}
Hence the total island density has scaling form 
\begin{equation}
\label{N_Scale1}
N \sim \theta^{1+z}
\end{equation}
As we will see, a naive analysis of the process gives $z=0$. If  $z\neq 0$ we say that
we have anomalous scaling.

The second quantity we are interested in is the island-island correlation function
$G(r)$, which is the  probability of finding two islands 
separated by a  distance $r$. Note that $r \leq 1 $ is inaccessible
in the model and we define  $G(0)=0$ here. For large distance $r$ the correlation
function should approach the total island density $N^2$ since there is no long range order
in the system.

The simulation results for $d=1$ dimension are summarized as follows. The simulation 
involved 1000 runs on lattices of  1000 sites, with periodic boundary 
conditions. We use critical size $i$ ranging from 2 to 4, and initial coverage $\theta$ ranging 
from 0.02 to 0.2. The results for $ N_s(\theta) $ are shown in Fig \ref{figNsd1}. 
A data collapse  shows 
that $z=0$ for $i=2$ and, $z=0.03$ for $i=3$, but $z=0.2$ for $i=4$. We will argue below
that this corresponds to anomalous scaling for $i \geq 4$. For $i=2$, $N_s$ can be compared
to an exact solution of a mean field rate equation: 
$N_s= e^{-1}\theta (s-1)/s! $ \cite{rateeq} 
which fits the simulation curve. We can 
see that the formation of islands mostly comes from the nucleation events: two
adatoms meet each other and become an island. The aggregation events are  
rare.  On the other hand, for large $i$, nucleation events are less probable than aggregation events.

The $z=0$ result can be understood by naive scaling theory, because there is only one 
characteristic length scale, and $N_s$ and $\theta$
should have the same dimension, $L^{-1}$, where $L$ is the system size. Suppose we
rescale the lattice constant: $a \rightarrow \lambda a$ $(\lambda >1)$ while keeping the number and distribution 
of the initial adatoms fixed. This  operation will reduce the initial 
coverage of adtoms to half its original value. Now, if we let the adtoms still 
perform the unit length hopping then it would be reasonable to assume that  
the final number of islands with 
size s will be the same as before. Thus $N_s$ is half of the previous value, 
that is $N_s \sim \theta$ or $ z=0 $.

The fact that $z \neq 0 $ for $i \geq 4$ implies that the {\it lattice constant}  
plays a relevant role in this case, 
since it is the only other length in the problem.
The exponent $z$ is analogous to  an anomalous dimension \cite{Goldenfeld} in  statistical physics. 
If we include the fact that $\theta$ has the dimension of $L^{-d}$ then the proper
scaling theory suggests that
\begin{equation}
\label{Ns_Scale3}
N_s(\theta)=a^{-2dz}\theta^{1+2z}f(s \theta^za^{dz})
\end{equation}
where $a$ is the lattice constant which has the dimension of length
 and $d$ is the dimension of the substrate. Similarly, the average island 
 size $S$ and total island density $N$ have the scaling form
\begin{equation}
\label{S_Scale2}
S \sim a^{dz}\theta^{-z}
\end{equation}
and 
\begin{equation}
\label{N_Scale2}
N \sim a^{dz}\theta^{1+z}
\end{equation}

One interesting property of this model is seen when $\theta \rightarrow 0$ for fixed
$a$, namely
$S \rightarrow \infty$ for $z>0$. It seems that there is 
 only one giant island finally for an initially sparsely distributed  system. 
This happens when  $i \geq 4$ in $d=1$.
 
We can attempt a qualitative explanation of our results. 
There are three fundamental physical processes in our model: diffusion,
nucleation and aggregation. The final island distribution is the result of local (esp. 
in $d=1$ dimension)  competition between nucleation and aggregation. It is well known 
that in $d=1$  
the  probability for  a particle to eventually return to its random walk origin is always $1$,
but the reunion of $i$ particles is not always certain. In fact, Fisher\cite{Fisher} showed that, in $d=1$, this
reunion has probability  1 when $i<4$ but smaller than $1$ when $i\geq4$.
We believe that this is reflected 
in our simulation in that once an island (for $i \geq 4$) is formed it will sweep up many 
adatoms. Adatoms have small probability to group in $4$ before they are
absorbed by an existing island.

The simulation for d=2 involves 1000 runs on a 100 times 100 lattice. The critical size
varies from 2 to 4 and the initial coverage from 0.02 to 0.2. The data for $N_s(\theta)$
are shown in (Fig \ref{figNsd2}). A data collapse  shows that
 $z=0$ for $i=2$ , $z=0.24$ for $i=3$ and 
$z=0.65$ for $i=4$. We have been able to extend Fisher's argument to $d=2$. We find 
that the probability for 3 walkers to meet is less than 1 in $d=2$. Our simulation 
indeed shows anomalous scaling for $i\geq 3$. 

Further, we studied the island-size distribution for $i=2$ in $d$ up to 4. 
Interestingly, we found that $z$ is always 0 and the island-size distribution function $N_s(\theta)$
agrees with the rate equation result 
	$N_s= e^{-1}\theta (s-1)/s!$\cite{rateeq}. 
The rate equation result is obtained under the assumption that the ``capture number'' of  an island is
independent of its size $s$ \cite{rateeq}. However, recently some authors\cite{PRL96Evans} 
pointed out that this ``capture number''
depends on  island size $s$ even in a point-island model. They simulated a point-island 
model with external flux of adatoms and obtained the ``capture number'' for each class  of  island
by measuring the capture event per unit time for each class of island.
They found 
that the capture number depends on the island size $s$ for large island while for small island
this number is a constant. Since we turn off flux of adatoms in our model
we do not have ( $i=2$ case ) enough adatoms to form  very large island. Thus the the constant
``capture number'' assumption is reasonable in our case. 

We have also studied the island-island spatial correlation function. 
The results are shown in Fig. \ref{figGr}. 
Starting from 0 at $r=0$ the island-island correlation function increases to an asymptotic
value which is the square of the total island density.  The  value of $G(r)$ near $r=0$ implies that
close to an island further nucleation is suppressed. Adatoms are
bound by the island edge before they meet another adatom. The 
depletion in the population of nearby pairs of islands is
enhanced with increasing $i$. This depletion leads to a ring structure or ``splitting" in
the diffraction profile of the specular beam\cite{JValSci}. For $i=2$ we get a functional 
form of $G(r)$
which in good agreement with   the  simulation. Our argument for $i=2$ is as follows.

Since an island is formed by an adatom and its nearest 
neighbor (essentially by definition for $i=2$ ) one can approximate
$G(r)$ (for islands) by  
\begin{equation}
  G(r) \sim  \sum_{n=2} ^{\infty} p_n(r)
\end{equation}
where $ p_n(r)$  the probability of initially (at time $t=0$) finding an n-th
nearest neighbour adtom at distance r from a given adatom. This probability is given 
by the  Poisson distribution for an randomly  distributed system,
\begin{equation}
p_n(r) \sim (r\theta)^n e^{-r\theta}/n!
\end{equation}
Therefore we get 
\begin{equation}
 G(r) \sim 1 - e^{-r\theta} .
\end{equation}
Generally we propose a closed form for G(r) 
\begin{equation}
\label{rojo}
 G(r) \sim  1-e^{-r\theta}-\dots -(r\theta)^{i-2}e^{-r\theta}/(i-2)! .
\end{equation}
From Fig\ref{figGrfit} we can see that this functional form of $G(r)$ fits the simulation for
$i$ up to 4 in $d=1$. 

The results for the island-island
spatial correlation function in $d=2$ are shown in Fig \ref{figGr}. For $i=3,4$ near 
$ r \sim 0 $ we only see the linear behavior of $G(r)$. This doesn't agree with the mean-field
theory we gave above . This can be understood by noting the  topological difference 
between $d=1$
dimension and  $d\geq 2$ dimension, namely in $d \geq 2$ dimension an adatom can go around
an island to meet another adatom. This behavior reduces the depletion effect of island in 
higher dimensions.

In our model we ignore the shape of the island. We expect that for small initial coverage the
shape of island will not affect the results we obtained, especially the universal index $z$ for 
the island size distribution. However, for large initial coverage we enter the percolation regime
where this assumption no longer holds. It will be interesting to study  a  more realistic model for 
a large (but still below the percolation critical value) initial coverage. 

Another interesting issue is the critical island size $i$. In our model we assumed that island 
with size larger than $i$ are stable and  immobile. In experiment the peripheral adatoms of an island
always have the chance to break the bond to the island. This will make the definition of {\it critical
island size} ambiguous. By changing the breaking energy we can effectively tune the critical island
size $i$ continously. It will be interesting to see when the anomalous behaviour appears.

In sumary, in this paper we studied a new point-island model in both 1 and 2 dimensions for various 
critical island size and different initial adatom coverage. We investigated the island size
distribution function for various initial coverages, critical island sizes and in different dimensions. 
We proposed a scaling form for the island size distribution function and  found anomalous 
behaviour for this  function when critical island size is large. This can be understood 
by the many body random walk theory, namely the reunion probability of $i$ random walkers is
smaller than 1 when $i$ is large enough. A search for experimental manifestation of this effect
would be interesting. We also studied island-island spatial correlation 
function. Using a version of mean field theory
we obtain a closed form for the island-island spatial correlation  function in 1 dimension. Monte-Carlo
simulation agrees with the analytic results. 

We would like to thank Ellak Somfai, Fred Mackintosh,  Christine Orme, and B.G. Orr 
for useful discussions.   JL and LMS are supported by NSF grant DMR 94-20335.

\pagebreak

\begin{figure}
\caption{simulation results of $N_s(\theta)$ and data collapse, $d=1$.
 We choose system size $L=10000$ and number of runs 1000. Here initial 
 coverage $\theta$ varies from 0.02 to 0.2. The circle symbol stands
 for $\theta =0.02$ and  square for $\theta =0.04$ etc.
\label{figNsd1}
}
\end{figure}

\begin{figure}
\caption{simulation results of $G(r)$ , $d=1$ and $d=2$.
$G(r)$ is normalized here so that $G(\infty)=1$. Here
initial coverage $\theta =0.1$ and number of runs 1000.
The system size $L$ is 10000 for $d=1$ and 100 for $d=2$
\label{figGr}
}
\end{figure}

\begin{figure}
\caption{fit of $G(r)$ using the functional form from mean field theory, $d=1$ .
Here initial coverage $\theta =0.1$ and number of runs 1000. 
The system size $L$ is 10000.
\label{figGrfit}
}
\end{figure}

\begin{figure}
\caption{simulation results of $N_s(\theta)$  and data collapse, $d=2$. 
We choose system size $L=100$ and number of runs 1000. Here initial 
 coverage $\theta$ varies from 0.02 to 0.2. The circle symbol stands
 for $\theta =0.02$ and  square for $\theta =0.04$ etc.\label{figNsd2}
}
\end{figure}

\end{document}